\documentclass[aps,eqsecnum,amsmath,amssymb,showpacs,nofootinbib]{revtex4}
\usepackage{graphicx}
\usepackage{dcolumn}
\usepackage{bm}

\begin{document} 
\title{\boldmath Tetraquark-adequate formulation of QCD sum rules}
\author{Wolfgang Lucha$^a$, Dmitri Melikhov$^{a,b,c}$, and Hagop
Sazdjian$^d$}
\affiliation{ $^a$Institute for High Energy Physics, Austrian
Academy of Sciences, Nikolsdorfergasse 18, A-1050 Vienna,
Austria\\ $^b$D.~V.~Skobeltsyn Institute of Nuclear Physics,
M.~V.~Lomonosov Moscow State University, 119991, Moscow, Russia\\
$^c$Faculty of Physics, University of Vienna, Boltzmanngasse 5,
A-1090 Vienna, Austria\\ $^d$Institut de Physique Nucl\'eaire,
CNRS-IN2P3, Universit\'e Paris-Sud, Universit\'e Paris-Saclay,
91406 Orsay, France
}
\date{\today}
\begin{abstract}
We study details of QCD sum rules \`a la
Shifman-Vainshtein-Zakharov for exotic tetraquark states. We point
out that duality relations for correlators involving exotic
currents have fundamental differences compared with the duality
relations for the correlators of bilinear quark currents: namely,
the $O(1)$ and $O(\alpha_s)$ [$\alpha_s$ the strong coupling
constant] terms in the operator product expansion for the exotic
correlators exactly cancel against the contributions of the
two-meson states on the hadron side of QCD sum rules. As a result,
the tetraquark properties turn out to be related to the specific
non-factorizable parts of the exotic Green functions; the relevant
non-factorizable diagrams start at order $O(\alpha_s^2)$.
\end{abstract}
\pacs{11.15.Pg, 12.38.Lg, 12.39.Mk, 14.40.Rt}
\maketitle

\section{Motivation} \label{s1}
Motivated by increasing experimental evidence for narrow
near-threshold hadron resonances with favorable interpretation as
tetraquark and pentaquark hadrons (of minimal parton
configurations consisting of four and five quarks, respectively)
\cite{Esposito:2016noz,ali,Olsen:2017bmm}, extensive theoretical
studies of such objects have been carried out. This letter focuses
on subtleties of the description of tetraquark mesons by
Shifman-Vainshtein-Zakharov (SVZ) sum rules in QCD \cite{QCD_SR};
we demonstrate that some essential criteria for selecting QCD
diagrams relevant for tetraquark properties within QCD sum rules
have not been properly taken into account.

For a proper QCD analysis of possible tetraquarks and for the
selection of the appropriate Feynman diagrams, the understanding
of the four-quark singularities of Feynman diagrams plays a
crucial role. In
Refs.~\cite{coleman,weinberg,knecht,cohen,maiani,lms_prd,lms_epjc,maiani2,lms_pos,lms_prd2018},
the four-quark singularities of Feynman diagrams describing the
four-point functions $\Gamma_{4j}$ of bilinear quark-antiquark
currents $j$ have been carefully studied. References
\cite{lms_pos,lms_prd2018} introduced the notion of
tetraquark-phile ($T$-phile) diagrams: the $T$-phile diagrams are
those Feynman diagrams that have four-quark singularities in the
relevant kinematic variable. For $\Gamma_{4j}$, those diagrams
that contain at least two gluon exchanges of special topology have
been shown to belong to the set of $T$-phile diagrams.

Independently of this line of research, numerous works deal with
the analysis of tetraquark states by QCD sum rules (see
\cite{QCD_SR_1,QCD_SR_2} and references therein). QCD sum rules
exploit dispersion representations to calculate QCD Green
functions, the time-ordered (T) products of local hadron
interpolating currents built of quark and gluon fields, in two
different ways: First, one calculates the Green function by
converting the T-product into a sum of local operators via
Wilson's operator product expansion (OPE). Power corrections,
reflecting the modification of quark and gluon propagators at
small momentum transfers due to QCD confinement, are given via QCD
condensates; they may be calculated, for each QCD diagram,
according to well-known rules \cite{QCD_SR}. In this way, one
obtains the sum rule's theoretical (OPE) side. Second, one
calculates the same Green function by inserting a complete set of
hadron states. This yields the sum rule's phenomenological
(hadronic) side.

For Green functions of currents $j$, the hadron continuum is
counterbalanced by the perturbative QCD contributions beyond an
appropriate effective threshold. Then, parameters of ordinary
hadrons are related to the low-energy region of perturbative QCD
diagrams supplemented by appropriate condensate contributions
\cite{QCD_SR}. We demonstrate that, for Green functions involving
tetraquark currents, this picture requires serious modifications.

References \cite{QCD_SR_1,QCD_SR_2} focused on two-point functions
$\Pi_{\theta \theta}(x)=\langle {\rm
T}\{\theta(x)\theta(0)\}\rangle$ of tetraquark interpolating
currents $\theta(x)=\bar q(x)q(x) \bar q(x) q(x)$ and three-point
functions $\Gamma_{\theta jj}(0|x,y)=\langle {\rm
T}\{\theta(0)j(x)j(y)\}\rangle$, involving one tetraquark current
$\theta$ and two ordinary currents $j(x)=\bar q(x)q(x)$; $\langle
...\rangle$ denotes averaging over the vacuum. (The currents'
quark flavor content will be specified below.) All previous
applications of SVZ sum rules (SR) to exotic states share one
common feature: they adopt the leading-order $O(1)$ diagrams (and
sometimes also radiative corrections) and power corrections
induced by these diagrams, and borrow exactly the same criteria
for continuum subtraction as prescribed for ordinary mesons
\cite{QCD_SR}. Consequently, the tetraquark contribution is found
to be dual to the low-energy spectral integral of the relevant QCD
diagrams. Specifically, tetraquarks receive substantial
contributions of $O(1)$ and $O(\alpha_s)$ QCD diagrams.

We will prove that the procedures adopted in SR analyses of exotic
states \cite{QCD_SR_1,QCD_SR_2} do not take proper account of the
cancellations between the $O(1)$ and $O(\alpha_s)$ diagrams on the
OPE side and the two-meson contributions on the hadron side. Let
us start with two almost self-evident observations:

\noindent (i) QCD sum rules utilize local interpolating currents,
so it suffices to consider tetraquark interpolating currents in
the form of products of two colorless bilinear quark currents
\cite{jaffe}. All other color structures of tetraquark currents
are reduced to products of colorless bilinears by Fierz
transformations. For the singlet-singlet color structure of
$\theta$, any diagram describing $\Pi_{\theta \theta}$ and
$\Gamma_{\theta jj}$ may be obtained from the diagrams of
$\Gamma_{4j}$ 
by merging two pairs of vertices (in the case of $\Pi_{\theta
\theta}$) or one pair of vertices (in the case of $\Gamma_{\theta
jj}$). Technically, the relationship between the Green functions
involving tetraquark currents and $\Gamma_{4j}$ corresponds to
defining the local $\theta(x)$ as the product of two point-split
colorless currents $j$ by sending their displacement $\delta$ to
zero: $\theta(x)=\lim\limits_{\delta\to
0}j(x)j(x+\delta)$.\footnote{The parameter $\delta$ finally sent
to zero must not be confused with the finite physical separation
between clusters inside tetraquarks discussed in dynamical models
for the tetraquark structure, e.g., \cite{brodsky,polosa}.}

\noindent (ii) All previous QCD SR applications to exotic states
\cite{QCD_SR_1,QCD_SR_2} relate tetraquark properties to those
contributions to $\Pi_{\theta \theta}$ and $\Gamma_{\theta jj}$
that are obtained by merging vertices in non-$T$-phile diagrams of
$\Gamma_{4j}$. Recall that such contributions to $\Gamma_{4j}$
have no four-quark cuts \cite{landau} and therefore may not be
related to tetraquark properties \cite{lms_prd,lms_epjc,maiani2}.
One may therefore doubt that the procedures adopted in
\cite{QCD_SR_1,QCD_SR_2} are consistent.

We will show that quark-hadron duality relations for Green
functions involving exotic tetraquark currents exhibit a specific
feature: an exact cancellation of the $O(1)$ and $O(\alpha_s)$
contributions on the OPE side against the two-meson contribution
on the hadron side of the SVZ sum rule by virtue of quark-hadron
duality relations for correlation functions of colorless currents
$j$. (This property is quite general and does not depend on the
color structure of $\theta$ but is most easily demonstrated for
$\theta$ taken as the product of two colorless $j$; we therefore
present here the analysis of this case.) Hence, upon taking into
account these cancellations a QCD SR for any exotic state assumes
the following form: the OPE side for $\Pi_{\theta\theta}$ and
$\Gamma_{\theta jj}$ has merely contributions of $T$-phile
diagrams, obtained from $T$-phile diagrams for $\Gamma_{4j}$ by
merging appropriate vertices; the hadron side has the suspected
tetraquark pole and the interacting mesons. One may then assume,
similarly to conventional QCD sum rules for ordinary correlators,
that the tetraquark contribution is dual to the low-energy part of
the $T$-phile contributions to $\Pi_{\theta\theta}$ and
$\Gamma_{\theta jj}$.

\section{Direct Green functions involving tetraquark currents}

Let us consider tetraquarks involving two quarks of flavors $a$
and $c$ and two antiquarks of flavors $b$ and $d$, and thus define
interpolating currents with two different flavor structures,
$\theta_{\bar a b\bar c d}=j_{\bar a b}j_{\bar c d}$ and
$\theta_{\bar a d\bar c b}=j_{\bar a d}j_{\bar c b}$, with
$j_{\bar ab}=\bar q_a q_b$. We need not specify the Dirac
structure of $\theta$, since it does not change our argument.

An appropriate definition of $\theta$ may be given by
point-splitting in the product of two currents $j$. From this
perspective, any diagram involving some $\theta$ may be obtained
from the four-point function of two currents $j$ studied in detail
in \cite{lms_epjc}. We distinguish between Feynman diagrams where
quark flavors in initial and final state are combined in the same
way (direct diagrams) and in a different way (recombination
diagrams), since they have different topologies and different
structures of four-quark singularities. Accordingly, the resulting
duality relations should be discussed separately.

\subsection{Two-point function $\Pi^{\rm dir}_{\theta \theta}$}
\begin{figure}[!t]
\includegraphics[width=5.5cm]{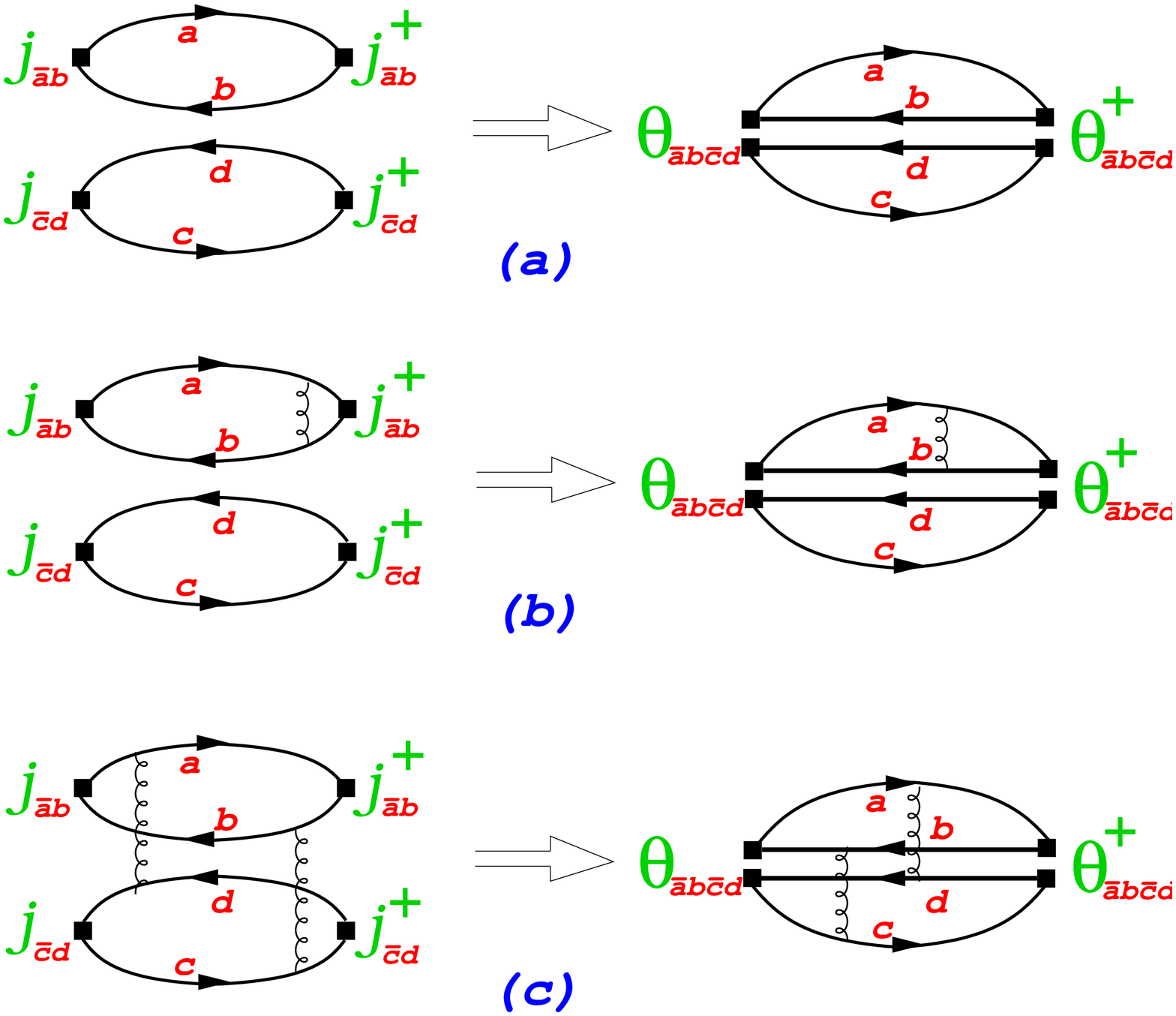}
\caption{\label{Fig:2pt_dir} Feynman diagrams for a direct
two-point function of tetraquark currents $\Pi_{\theta\theta}^{\rm
dir}$, obtained by merging vertices in $\Gamma^{\rm dir}_{4j}$. In
the left column, diagrams (a) and (b) do not contain four-quark
singularities in the $s$ channel, whereas diagram (c) is the
lowest-order diagram that contains the four-quark $s$ cut and is
thus the only $T$-phile diagram. }
\includegraphics[width=11cm]{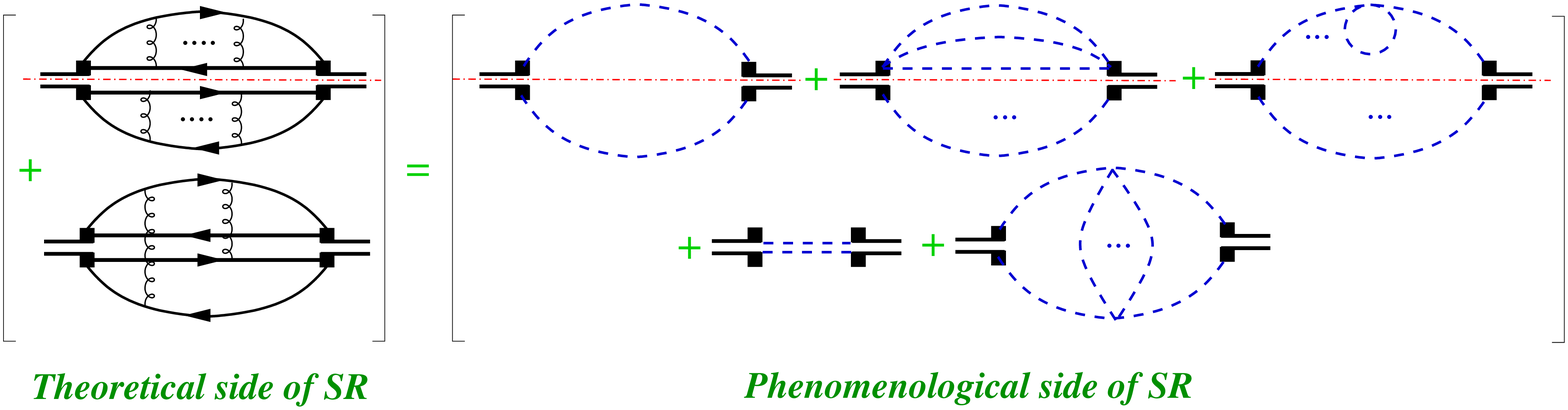}
\caption{\label{Fig:esr1} QCD SR for the two-point function
$\Pi^{\rm dir}_{\theta\theta}$: The l.h.s.\ shows its OPE.
The r.h.s.\ shows its meson representation; assuming that the
hadron spectrum contains a tetraquark, its contribution appears on
the hadron side. The first line on both sides of the SR shows
diagrams factorizable into two parts, separated by the red
dash-dotted line; the second line on both sides shows
non-factorizable contributions. NB. the set of factorizable
diagrams on the OPE side includes diagrams (a) and (b) of the
r.h.s. of Fig.~\ref{Fig:2pt_dir}. Diagrams in the first line on
both sides of the SR are equal to each other by virtue of QCD sum
rules for $\Pi_{jj}$, Fig.~\ref{Fig:esr2}.} \vspace{.4cm}
\includegraphics[width=9cm]{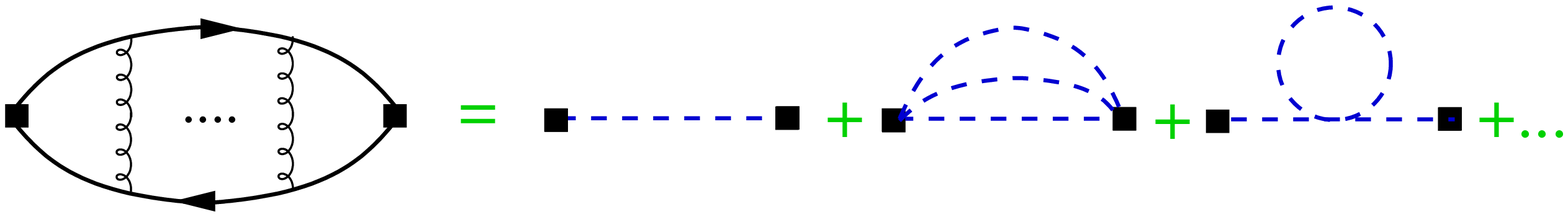}
\vspace{-.3cm} \caption{\label{Fig:esr2} Conventional QCD SR for
the two-point function $\Pi^{\rm dir}_{jj}$ of ordinary currents
$j$: the l.h.s.\ shows its OPE (the diagram with two gluon lines
with dots in-between represents the sum of diagrams with an
arbitrary number of gluon exchanges, starting with the quark loop
with no gluons); the r.h.s.\ shows its meson representation.}
\vspace{.4cm}
\includegraphics[width=9cm]{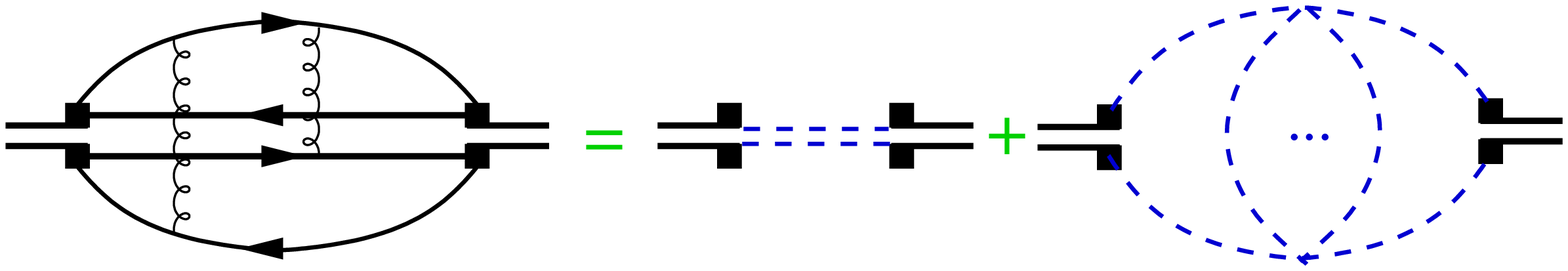}
\vspace{-.3cm} \caption{\label{Fig:esr3} Final tetraquark QCD SR
relating the non-factorizable OPE contributions to the two-point
function $\Pi_{\theta\theta}^{\rm dir}$ to the sum of the
tetraquark pole and the non-factorizable meson-interaction
diagrams. The dots in the meson diagrams denote the sum of meson
diagrams of the same (non-factorizable) topology.}
\end{figure}

Figure~\ref{Fig:2pt_dir} shows the direct four-point function
$\Gamma^{\rm dir}_{4j}$ and the corresponding two-point function
of tetraquark currents: only diagrams (c) are $T$-phile, so the
r.h.s.\ diagrams (a,b) should drop out from the tetraquark SR.
Diagrams with one-gluon exchanges between disconnected loops are
null.

To show that this indeed happens, we inspect the OPE and the
hadron representation for the two-point direct correlation
function $\Pi_{\theta\theta}$, Fig.~\ref{Fig:esr1}. [In
Figs.~\ref{Fig:esr1}-\ref{Fig:esr3}, we do not explicitly show
power corrections: for any Feynman diagram, they are calculated
according to well-known rules \cite{QCD_SR}.] There is an infinite
subset of diagrams in the OPE for $\Pi^{\rm dir}_{\theta\theta}$
that factorize in coordinate space into two parts separated by the
red dash-dotted line; the $O(1)$ and $O(\alpha_s)$ diagrams belong
to this subset. On the hadron side, there is also an infinite
subset of meson contributions that factorize in coordinate space.
It is straightforward to check that the OPE factorizable subset
exactly equals the hadron factorizable subset by using the QCD SR
for the two-point function $\Pi_{jj}$ of ordinary currents $j$
(Fig.~\ref{Fig:esr2}). Cancelling out the equal factorizable parts
on both sides of the SR of Fig.~\ref{Fig:esr1}, we arrive at the
tetraquark SR of Fig.~\ref{Fig:esr3}. Now, similarly to the case
of ordinary mesons, we consider a single spectral representation
in $p^2$ of the QCD diagrams in the l.h.s.\ of Fig.~\ref{Fig:esr3}
and introduce an effective threshold $s_{\rm eff}$
\cite{lms_seff1,lms_seff2,lms_seff3} such that the contribution of
the QCD diagrams in the l.h.s.\ of Fig.~\ref{Fig:esr3} above
$s_{\rm eff}$ cancels the non-factorizable meson-meson interaction
diagrams on the r.h.s.\ of Fig.~\ref{Fig:esr3}. Then, after Borel
transformation from $p^2$ to Borel variable $\tau$ \cite{QCD_SR},
we obtain the ultimate tetraquark SR
\begin{eqnarray}
\label{2.1} (f_T^{\bar ab\bar cd})^2
\exp(-M_T^2\tau)=\int_{(4m_q)^2}^{s_{\rm eff}} ds \exp(-s
\tau)\rho^{\rm dir}_{T}(s)+\mbox{power corrections}.
\end{eqnarray}
Here, $4m_q\equiv m_a+m_b+m_c+m_d$, $\rho^{\rm dir}_{T}$ is the
spectral density in the variable $s$ of the r.h.s.\ of
Fig.~\ref{Fig:2pt_dir}(c) with two-gluon exchanges of order
$O(\alpha_s^2)$. Power corrections in (\ref{2.1}) correspond to
condensate insertions in the diagram of Fig.~\ref{Fig:2pt_dir}(c).
Power corrections generated by the r.h.s.\ diagrams in
Figs.~\ref{Fig:2pt_dir}(a,b), do not contribute to the tetraquark
SR (\ref{2.1}): they cancel against the factorizable meson-meson
contributions. $M_T$ is the tetraquark mass and $f^{\bar ab\bar
cd}_T=\langle T|\theta_{\bar ab\bar cd}|0\rangle.$ Only the
$T$-phile diagram of Fig.~\ref{Fig:2pt_dir}(c) and the
corresponding power corrections contribute to the tetraquark SR
(\ref{2.1}).

${}$
\subsection{Three-point function $\Gamma^{\rm dir}_{\theta jj}$}
\begin{figure}[!ht]
\includegraphics[width=6.0cm]{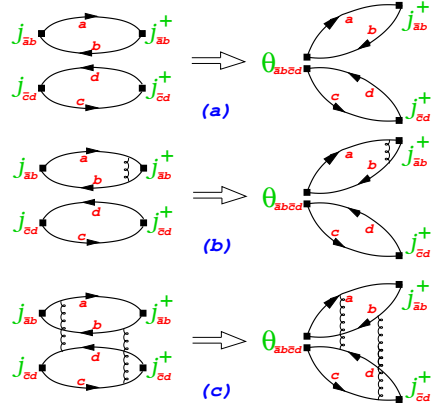}
\caption{\label{Fig:3pt_dir} Feynman diagrams for a direct
three-point function as obtained by merging vertices in
$\Gamma^{\rm dir}_{4j}$. Diagrams (a,b) depend on $p^2$ at most
polynomially and therefore cannot contribute to tetraquark
properties. Only diagrams (c) are $T$-phile and therefore
contribute to the tetraquark SVZ SR.}
\end{figure}

Direct Green functions $\Gamma^{\rm dir}_{\theta jj}$ may be found
from $\Gamma^{\rm dir}_{4j}$ by merging in the latter just one
(say, the left) coordinate pair, as shown in Fig.~\ref{Fig:3pt_dir}; here, only diagram (c) is
$T$-phile, so the r.h.s.\ diagrams (a) and (b) should not
contribute to the tetraquark coupling to two mesons. For
$\Gamma^{\rm dir}_{\theta jj}$, this is easily shown: the
tetraquark would lead to a pole $1/(p^2-M_T^2)$ in $\Gamma^{\rm
dir}_{\theta jj}$ ($p$ is the total momentum of the currents) with
residue related to the tetraquark's coupling to ordinary mesons of
appropriate flavor content, $T\to M_{\bar a b}M_{\bar c d}$.
Clearly, the diagrams of Figs.~\ref{Fig:3pt_dir}(a,b) cannot
contribute to the pole, as their dependence on $p^2$ is at most
polynomial, due to traces over quark loops. The Borel transform of
the r.h.s.\ of Figs.~\ref{Fig:3pt_dir}(a,b) vanishes. So the
r.h.s.\ diagram of Fig.~\ref{Fig:3pt_dir}(c) is the lowest-order
diagram that gives a nontrivial contribution to the tetraquark
pole. Introducing an effective threshold $s_{\rm eff}$ and
performing the Borel transform yields the SR
\begin{eqnarray}
\label{2B.1} f^{\bar ab \bar cd}_T\exp(-M_T^2\tau)A(T\to j_{\bar
ab}j_{\bar cd})=\int_{(4m_q)^2}^{s_{\rm eff}} ds \exp(-s
\tau)\Delta^{\rm dir}_{T}(s)+\mbox{power corrections}.
\end{eqnarray}
Here, $A(T\to j_{\bar ab}j_{\bar cd})$ is the momentum-space
amplitude $\langle 0|{\rm T}\{ j_{\bar ab}(x) j_{\bar
cd}(0)\}|T(p)\rangle$ and $\Delta^{\rm dir}_{T}(s)$ the spectral
density in the variable $s$ of the r.h.s.\ of
Fig.~\ref{Fig:3pt_dir}(c). As before, power corrections generated
by non-$T$-phile diagrams do not appear in the tetraquark SR.


${}$

\section{Recombination Green functions involving tetraquark currents
\label{3}} 
\begin{figure}[!ht]
\includegraphics[height=5.5cm]{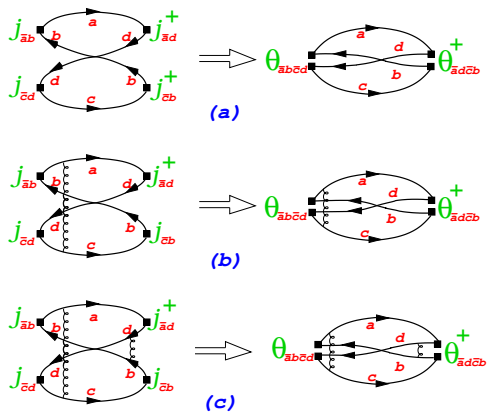}
\caption{\label{Fig:2pt_rec} Feynman diagrams for a recombination
two-point function of tetraquark currents as obtained by merging
vertices in the recombination four-point function of bilinear
quark currents. Diagrams (a) and (b) on the l.h.s.\ do not contain
four-quark singularities in the $s$ channel. Diagram (c) on the
l.h.s.\ is the lowest-order diagram that contains a four-quark $s$
cut; thus, only diagram (c) on the l.h.s.\ is the $T$-phile
diagram. Also, among the diagrams on the r.h.s., only diagram (c)
contributes to the OPE side of the tetraquark SVZ sum rule
(\ref{3.1}).}
\end{figure}

For correlators with recombination topology
(Fig.~\ref{Fig:2pt_rec}), where the initial and final quark color
singlets have different flavor structures, again only the diagrams
of Fig.~\ref{Fig:2pt_rec}(c) are $T$-phile and contribute to the
tetraquark SR. The proof of this is not as straightforward as
before. However, in \cite{lms_epjc} its was shown that the
recombination Green functions (a,b) on the l.h.s.\ of
Fig.~\ref{Fig:2pt_rec} do not contribute to tetraquark poles and
are related to meson amplitudes without $s$-channel four-quark
singularities. This holds even after merging initial and/or final
vertices. Thus, the r.h.s.\ diagrams (a,b) of
Fig.~\ref{Fig:2pt_rec} still do not contribute to the tetraquark
pole: only the diagrams in Fig.~\ref{Fig:2pt_rec}(c) are
$T$-phile. Similar considerations apply, mutatis mutandis, to the
recombination three-point functions. Hence, we arrive at the
tetraquark sum rules
\begin{eqnarray}
\label{3.1} f^{\bar ab\bar cd}_Tf^{\bar ad\bar
cb}_T\exp(-M_T^2\tau)&=&\int_{(4m_q)^2}^{s_{\rm eff}} ds \exp(-s
\tau)\rho^{\rm rec}_{T}(s)+\mbox{power corrections},\\ \label{3.2}
f^{\bar ab\bar cd}_T A(T\to j_{\bar ad}j_{\bar
cb})\exp(-M_T^2\tau)&=&\int_{(4m_q)^2}^{s_{\rm eff}} ds \exp(-s
\tau)\Delta^{\rm rec}_{T}(s)+\mbox{power corrections}.
\end{eqnarray}
Here, $4m_q\equiv m_a+m_b+m_c+m_d$ and $\rho^{\rm rec}_{T}(s)$ and
$\Delta^{\rm rec}_{T}(s)$ are the spectral densities in the
variable $s$ of the $O(\alpha_s^2)$ diagrams with two-gluon
exchanges [cf.\ the r.h.s.\ of Fig.~\ref{Fig:2pt_rec}(c)]. The
coupling constants are defined by $f^{\bar ab\bar cd}_T=\langle
T|\theta_{\bar ab\bar cd}|0\rangle$ and $ f^{\bar ad\bar
cb}_T=\langle T|\theta_{\bar ad\bar cb}|0\rangle$.

\vspace{-.5cm}
\section{Conclusions}
\label{c}

We scrutinized the derivation of SVZ sum rules for exotic
correlation functions of tetraquark currents $\theta$, namely,
two-point functions $\Pi_{\theta\theta}$ and three-point functions
$\Gamma_{\theta jj}$ involving one tetraquark current and two
bilinear quark currents~$j$. Our insights may be summarized as
follows:
\begin{itemize}
\item[(i)]
The duality relations for exotic correlators are fundamentally
different from those for correlators of bilinear quark currents:
the $O(1)$ and $O(\alpha_s)$ terms
in the OPE for exotic correlators exactly cancel against the
contributions of two-meson states on the hadron side of QCD SR.
Thus, the properly formulated tetraquark SVZ sum rules,
Eqs.~(\ref{2.1}), (\ref{2B.1}), (\ref{3.1}), and (\ref{3.2}),
relate tetraquark properties to specific $T$-phile
non-factorizable parts of the OPE for exotic Green functions; the
corresponding non-factorizable diagrams start at order
$O(\alpha_s^2)$.

This result is quite general and valid for any color structure of
$\theta$, even if of diquark-antidiquark form. Merely the proof of
this statement becomes technically more involved: start with the
momentum-space diagrams generated by the diquark-antidiquark,
perform Fierz transformations reducing the diagrams to the ones of
singlet-singlet currents, and only then exploit the ordinary QCD
sum rules for the two-point functions of color-singlet currents.
Since Fierz transformations do not change a diagram's perturbative
order, one again observes the cancellation of the $O(1)$,
$O(\alpha_s)$, and the factorizable part of the $O(\alpha_s^2)$
contributions on the OPE side against the factorizable two-meson
contributions on the hadron side.

\item[(ii)]
For clarity, we demonstrated the aforementioned general property
by considering currents $\theta$ involving quarks of four
different flavors, which case exhibits the simplest topology of
the direct Green function. If some of the quark flavors in
$\theta$ coincide, the direct Green functions receive
contributions similar to those of their recombination
counterparts. The cancellation of the $O(1)$, $O(\alpha_s)$, and
the factorizable part of the $O(\alpha_s^2)$ corrections on the
OPE side against the factorizable two-meson contributions on the
hadron side of QCD sum rules becomes technically more involved but
has been verified.

\item[(iii)]
The cancellation of $O(1)$, $O(\alpha_s)$, and factorizable part
of the $O(\alpha_s^2)$ corrections in $T$-adequate sum rules holds
independently of the color structure of $\theta$. The
singlet-singlet color structure, however, has a decisive advantage
related to the algorithm of selecting $T$-phile $O(\alpha_s^2)$
diagrams: for $\theta$ chosen as the product of two color
singlets, the set of $T$-phile diagrams contributing to the
properties of some exotic state involves only diagrams obtainable
from $T$-phile diagrams of four-point functions of ordinary
currents $j$ by merging the appropriate vertices.
This observation reduces the analysis of duality relations for
tetraquark correlation functions to the analysis of four-quark
singularities in four-point functions of currents $j$. For other
color structures of $\theta$, the selection criteria for the
$T$-phile $O(\alpha_s^2)$ diagrams cannot be formulated in such a
direct manner: the $T$-phile diagrams for other color structures
of $\theta$ are just the Fierz-transformed $T$-phile diagrams
established for the singlet-singlet color structure of $\theta$.
\end{itemize}

The proper application of tetraquark QCD SR requires the knowledge
of presently unknown non-factorizable $O(\alpha_s^2)$ radiative
corrections and calls for further efforts in order to obtain
reliable conclusions about tetraquark candidates.

\vspace{.5cm} \noindent{\it{\bf Acknowledgements.}} The authors
are grateful to V.~Anisovich, T.~Cohen, L.~Gladilin, F.-K.~Guo,
M.~Knecht, L.~Maiani, B.~Moussallam, A.~Polosa, V.~Riquer,
S.~Simula, B.~Stech, and W. Wang for valuable discussions.
D.~M.~acknowledges support from the Austrian Science Fund (FWF),
project~P29028. D.~M. and H.~S. are grateful for support under
joint CNRS/RFBR grant PRC Russia/19-52-15022. D.~M. has pleasure
to thank L.~Maiani and W.~Wang for hospitality during his visit to
the T.~D.~Lee Institute.

\end{document}